\documentclass[prb,twocolumn,showpacs,preprintnumbers,amsmath,amssymb]{revtex4-1}

\usepackage{graphicx}
\usepackage{dcolumn}
\usepackage{bm}

\begin{document}

\title{Electronic structure, disconnected Fermi surfaces and antiferromagnetism
in the layered pnictide superconductor Na$_x$Ba$_{1-x}$Ti$_2$Sb$_2$O}

\author{David J. Singh}

\affiliation{Materials Science and Technology Division,
Oak Ridge National Laboratory, Oak Ridge, Tennessee 37831-6056}

\date{\today}

\begin{abstract}
We report electronic structure calculations for BaTi$_2$Sb$_2$O
and discuss the results in relation to the observed superconductivity
of this material when hole doped with Na.
The Fermi surface shows several sheets. These
include a nested
nearly 2D cylinder. There are also two
sheets, which are three dimensional, in spite of the layered crystal structure.
A magnetic instability associated with Fermi surface nesting is
found. A sign-changing $s$-wave state, different from the
one in the Fe-based superconductors, is predicted within a scenario
of spin-fluctuation mediated superconductivity.
\end{abstract}


\maketitle

\begin{figure}
\includegraphics[width=\columnwidth]{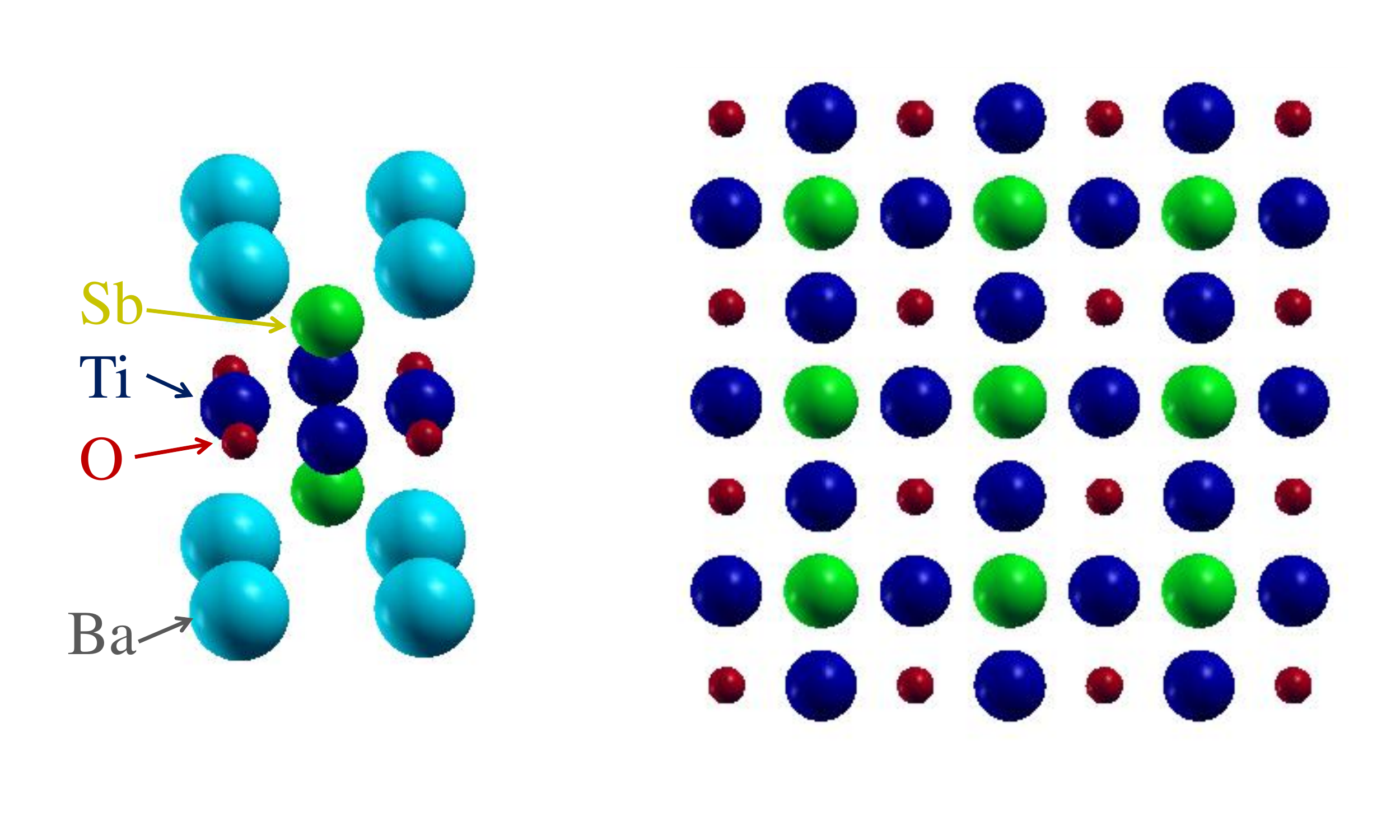}
\caption{Layered structure of BaTi$_2$Sb$_2$O (left) and as viewed
along the $c$-axis with the Ba spheres suppressed (right).}
\label{structure}
\end{figure}

The discovery of high temperature
superconductivity in Fe-based compounds
\cite{kamihara}
has lead to renewed interest in non-oxide superconductors.
Recently,
Doan and co-workers \cite{doan} reported superconductivity
in Ba$_{1-x}$Na$_x$Ti$_2$Sb$_2$O, 0.05$\leq$$x$$\leq$0.33
and $T_c$ up to 5.5 K.
The structure of the parent compound, BaTi$_2$Sb$_2$O,
as determined by x-ray refinement, \cite{doan}
is shown in Fig. \ref{structure}.
This structure shares a number of common features with the
Fe-based superconductors.
Specifically, it is a layered structure containing square
planes of transition element atoms, in this case Ti.
Like the Fe-based superconductors, there are two transition metal
atoms per chemical unit cell.
As may be seen, the two Ti in the BaTi$_2$Sb$_2$O unit cell
are related by a translation of [1/2,1/2] followed by a 90$^\circ$
rotation, and not by the glide of the Fe$_2$As$_2$ layers
in the Fe-based superconductors.
Superconductivity was also very recently reported by Yajima and
co-workers in undoped BaTi$_2$Sb$_2$O with a bulk
$T_c$ of 0.8 K. \cite{yajima}

The nearest neighbor Ti-Ti distance is 2.91 \AA,
which is only somewhat larger than the nearest
neighbor distance of 2.51 \AA{} in Ti metal,
implying the possibility of important direct Ti-Ti
bonding, as for the Fe lattice in Fe-based superconductors. \cite{singh-du}
Also similar to the Fe-based superconductors each Ti atom is coordinated
by four pnictogens, although in the present case the geometry is that
of approximate square planes with Ti at the center, instead of 
approximate tetrahedra and furthermore, no Fe-based superconductor is
known where the pnictogen is Sb. Furthermore, the Ti are also coordinated
by O in the present compound. Each Ti has two O nearest neighbors.
Thus as shown in the right panel of Fig. \ref{structure},
if one considers only Ti-O hopping, one obtains a bipartite
lattice consisting of two independent sets of one dimensional ...Ti-O-Ti-O...
chains, running respectively along $x$ and $y$.
Similarly if one considers only Ti-Sb hopping one also obtains one dimensional
chains, ...Ti-Sb$_2$-Ti-Sb$_2$..., running along $y$ and $x$ where Sb$_2$
denotes an Sb above the plane and an Sb below the plane. Such 1D objects
might therefore form the basis of a strongly nested electronic structure.
However, as mentioned, the short Ti-Ti nearest neighbor distance suggests
metal-metal bonding, which would couple the two sub-lattices.
Furthermore the Sb atoms are arranged in linear chains along the $c$-axis
with Sb-Sb distances of $\sim$$c$/2=4.05 \AA{}, which is short enough
that $c$-axis hopping could be important, especially assisted by the Ba.
Returning to the metal-metal bonding, 
one sees that the bonds run along the diagonal
directions in the two Ti atom unit cells, i.e. forming a square
network oriented at 45$^\circ$ to the ...Ti-O-Ti-O... and
...Ti-Sb$_2$-Ti-Sb$_2$... chains.

Experiments \cite{doan}
show that stoichiometric BaTi$_2$Nb$_2$O has a transition
at $T_s$=54 K, similar to several other related compounds.
\cite{axtell,liu2,ozawa,ozawa2,ozawa-stam}
This transition has strong signatures in resistivity, susceptibility
and specific heat. The microscopic mechanism for this transition
has not been demonstrated, but it is described as either a
charge density wave (CDW) or a spin density wave (SDW).
This is consistent with electronic structure calculations
for the compound Na$_2$Ti$_2$Sb$_2$O, which show a nested
Fermi surface. \cite{pickett}

In any case, as Na is substituted for Ba in
Na$_x$Ba$_{1-x}$Ti$_2$Sb$_2$O, $T_s$ is systematically
suppressed and superconductivity arises with $T_c$ of
up to 5.5 K. \cite{doan}
Importantly, superconductivity co-exists with the phase
corresponding to $T_s$ including near the maximum $T_c$.
Superconductivity occurs for 0.05$\leq$$x$$\leq$0.33,
which nominally amounts to average Ti valences between Ti$^{3.025+}$ and
Ti$^{3.165+}$ assuming that Sb occurs as Sb$^{3-}$. Titanates
with valences between 4+ and 3+ are known to have electron-phonon
superconductivity as in the case of electron
doped SrTiO$_3$, \cite{koonce}
but they can also be magnetic and exhibit non-trivial correlation
effects as in e.g. TiOCl and LaTiO$_3$.
\cite{saha,khaliullin}

Here we report first principles studies of BaTi$_2$Sb$_2$O
in relation to magnetism and superconductivity. We find that
like its sister compound Na$_2$Ti$_2$Sb$_2$O,
the Fermi surface is nested. This nesting leads to peaks in the
susceptibility that are incommensurate
but near the $X$ points. These peaks are strong enough
to result in an actual magnetic instability, which we confirm
by direct calculations. This shows that $T_s$ is most likely
an SDW instability. Therefore the superconductivity of doped
BaTi$_2$Sb$_2$O is in proximity to an antiferromagnetic SDW
phase. This leads to a specific prediction of the pairing
state in a spin-fluctuation mediated scenario, in particular
a sign changing $s$-wave state, but not the same one as in the
Fe-based superconductors. \cite{mazin-spm}

The electronic structure was calculated using standard density
functional theory based on the Perdew, Burke, Ernzerhof generalized
gradient approximation. \cite{pbe}.
For this purpose we used the general potential linearized
augmented planewave method including local orbitals \cite{singh-book}
as implemented in the WIEN2k code. \cite{wien2k}
The LAPW sphere radii were 2.4 bohr for Ba and Sb, 2.15 bohr for
Ti and 1.7 bohr for O. The calculations are based on the experimental
crystal structure.\cite{doan} This structure contains one
internal coordinate, $z_{\rm Sb}$=0.2514
corresponding to the Sb height.
The value obtained from total energy minimization of $z_{\rm Sb}$=0.2522
is very close to the experimental value.
This is in contrast to the large discrepancies found in the Fe-based
superconductors in standard non-magnetic
density functional calculations. \cite{mazin-mag}
We used well converged LAPW basis sets, including additional local orbitals,
along with dense Brillouin zone samplings, as high as a 48x48x24
mesh for the primitive tetragonal cell.

\begin{figure}
\includegraphics[width=0.9\columnwidth]{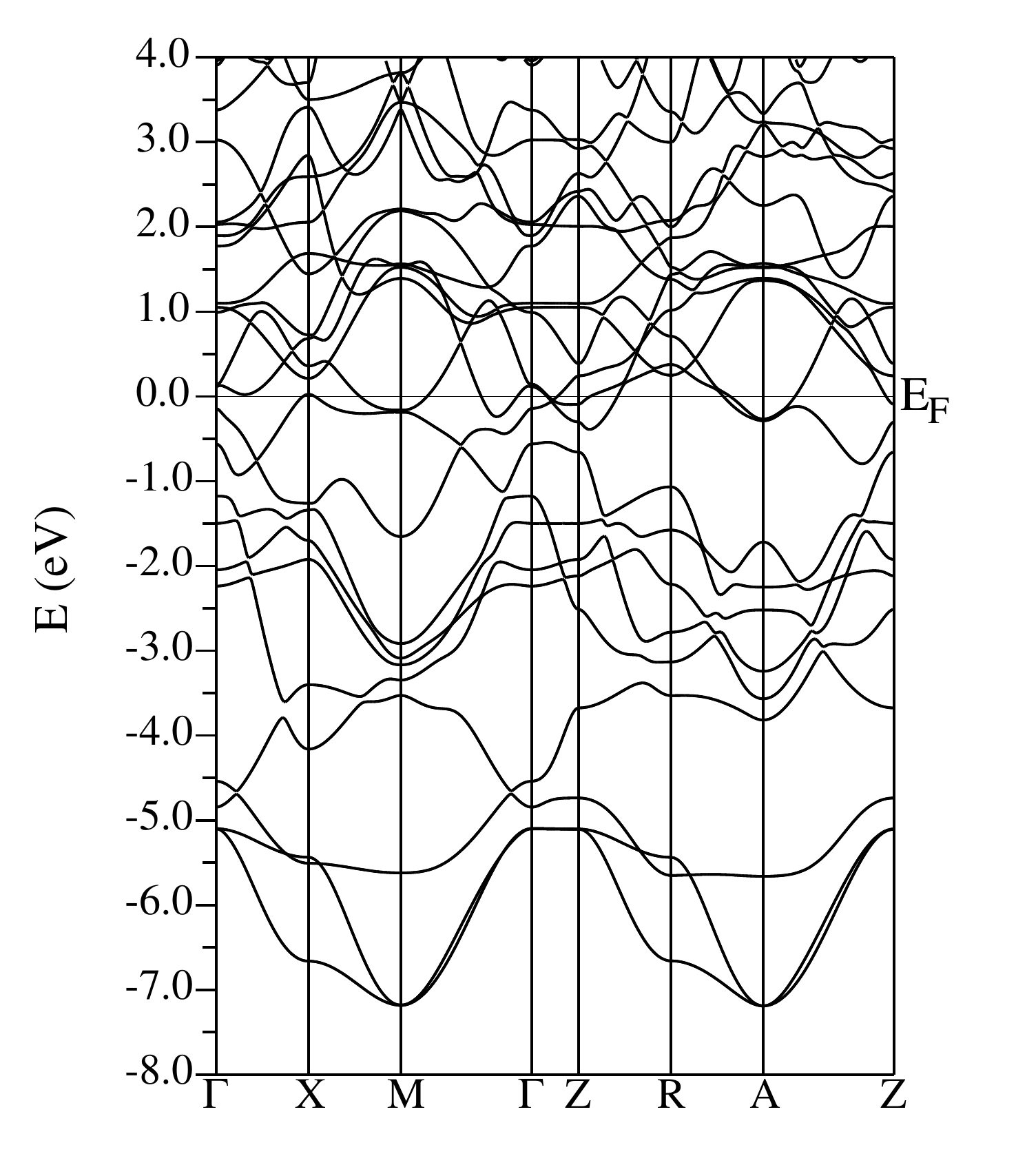}
\caption{Calculated band structure of BaTi$_2$Sb$_2$O including
spin orbit. The Fermi energy is denoted by the horizontal
line at 0 eV.}
\label{bands}
\end{figure}

\begin{figure}
\includegraphics[width=0.9\columnwidth,angle=0]{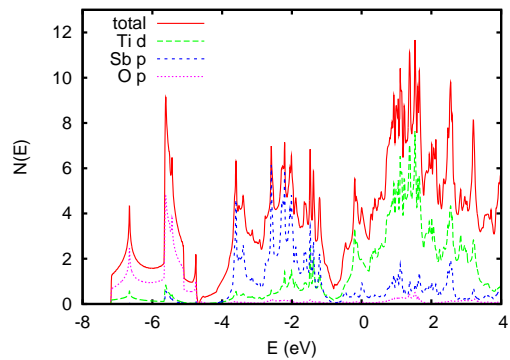}
\caption{Electronic density of states and projections onto
the LAPW spheres. }
\label{dos}
\end{figure}

\begin{figure}
\includegraphics[width=0.9\columnwidth,angle=0]{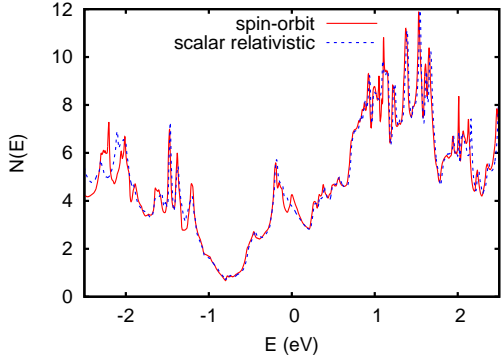}
\caption{Comparison of the density of states around $E_F$ with and without
spin orbit.}
\label{sr}
\end{figure}

The calculated band structure is shown in Fig. \ref{bands}
and the corresponding electronic density of states (DOS) is shown
in Fig. \ref{dos} along with projections.
These calculations were done relativistically, including spin-orbit,
but spin-orbit does not have a large effect.
It does, however, affect the details at $E_F$.
A comparison of the DOS with and without spin-orbit is given in Fig. \ref{sr}.

\begin{figure}
\includegraphics[width=0.90\columnwidth]{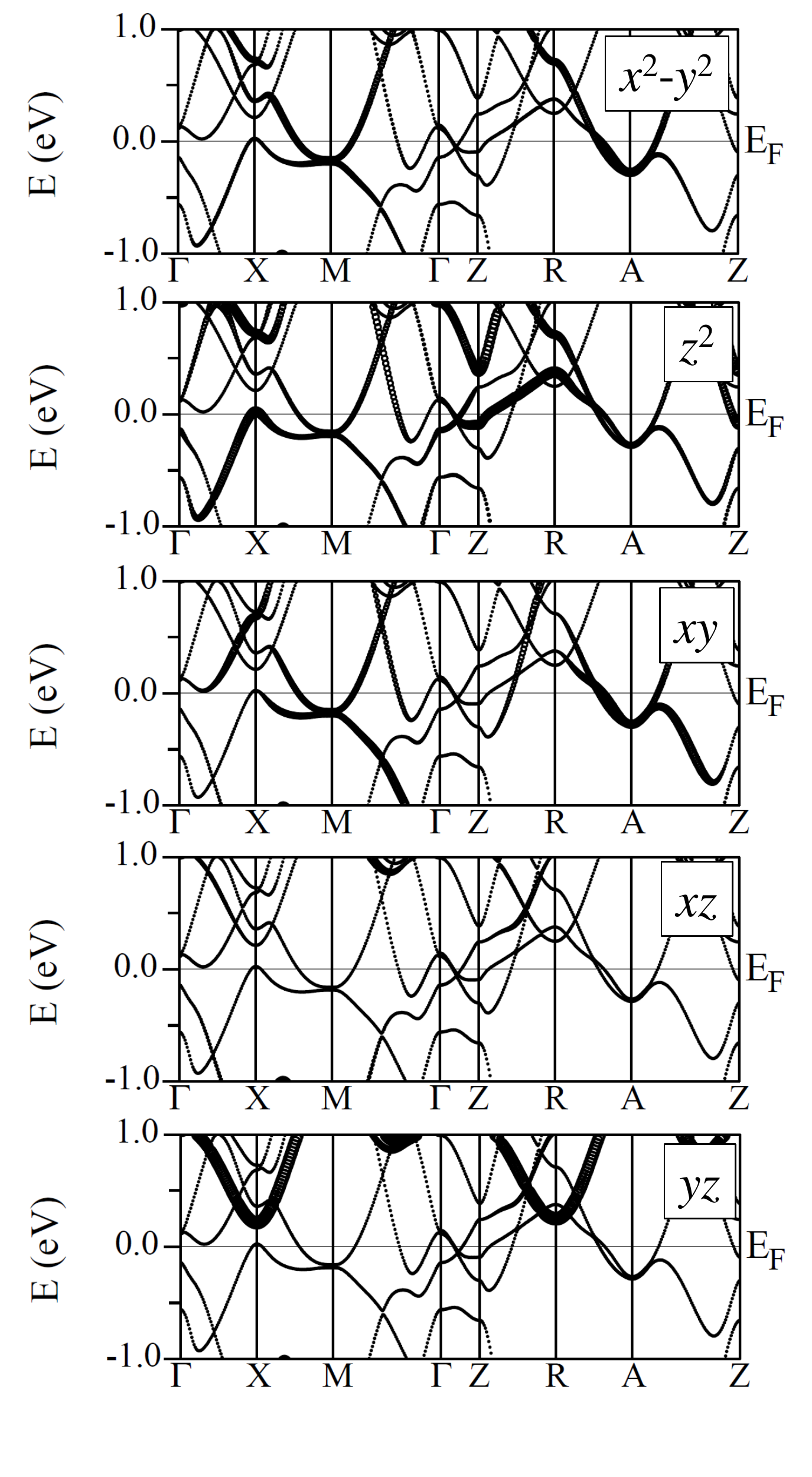}
\caption{Fat bands plot for different $d$ orbital characters
near $E_F$. The coordinate system is such that $d_{z^2}$ is directed
along the $c$-axis, and $x$ and $y$ are along the $a$-axis and $b$
axis, such that $d_{xz}$ is towards the nearest Sb and $y$ is in
the direction of the O neighbors.}
\label{fatbands}
\end{figure}

\begin{figure}
\includegraphics[width=\columnwidth]{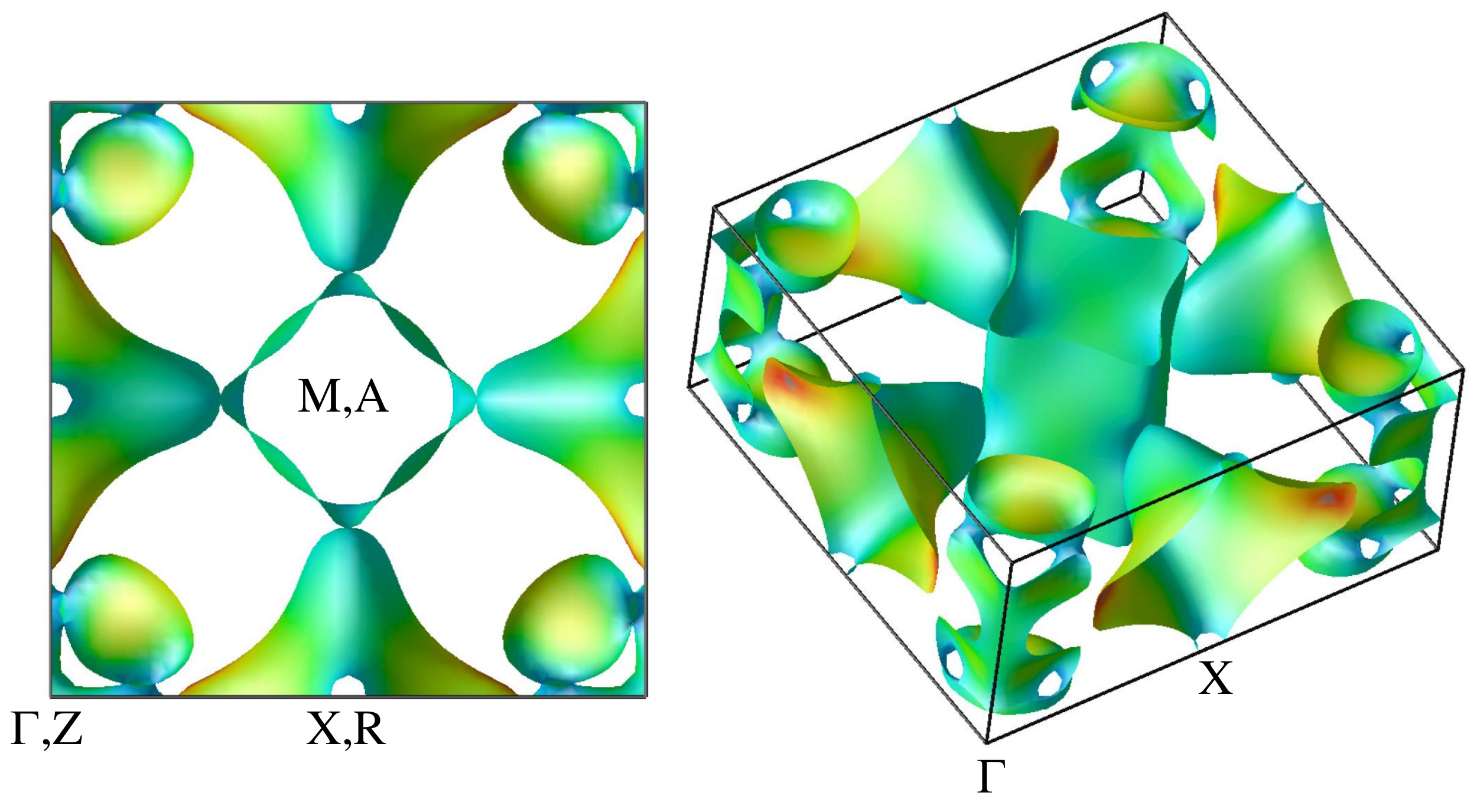}
\caption{Calculated Fermi surfaces of BaTi$_2$Sb$_2$O
spin orbit. The shading is by velocity, with blue
representing low velocity. The approximately square
cylindrical sections around the zone corner (M,A) are
electron sections as is the complex shaped section around
$\Gamma$ and Z. The other section around R is a
hole section.}
\label{fermi}
\end{figure}

\begin{figure}
\includegraphics[width=0.9\columnwidth]{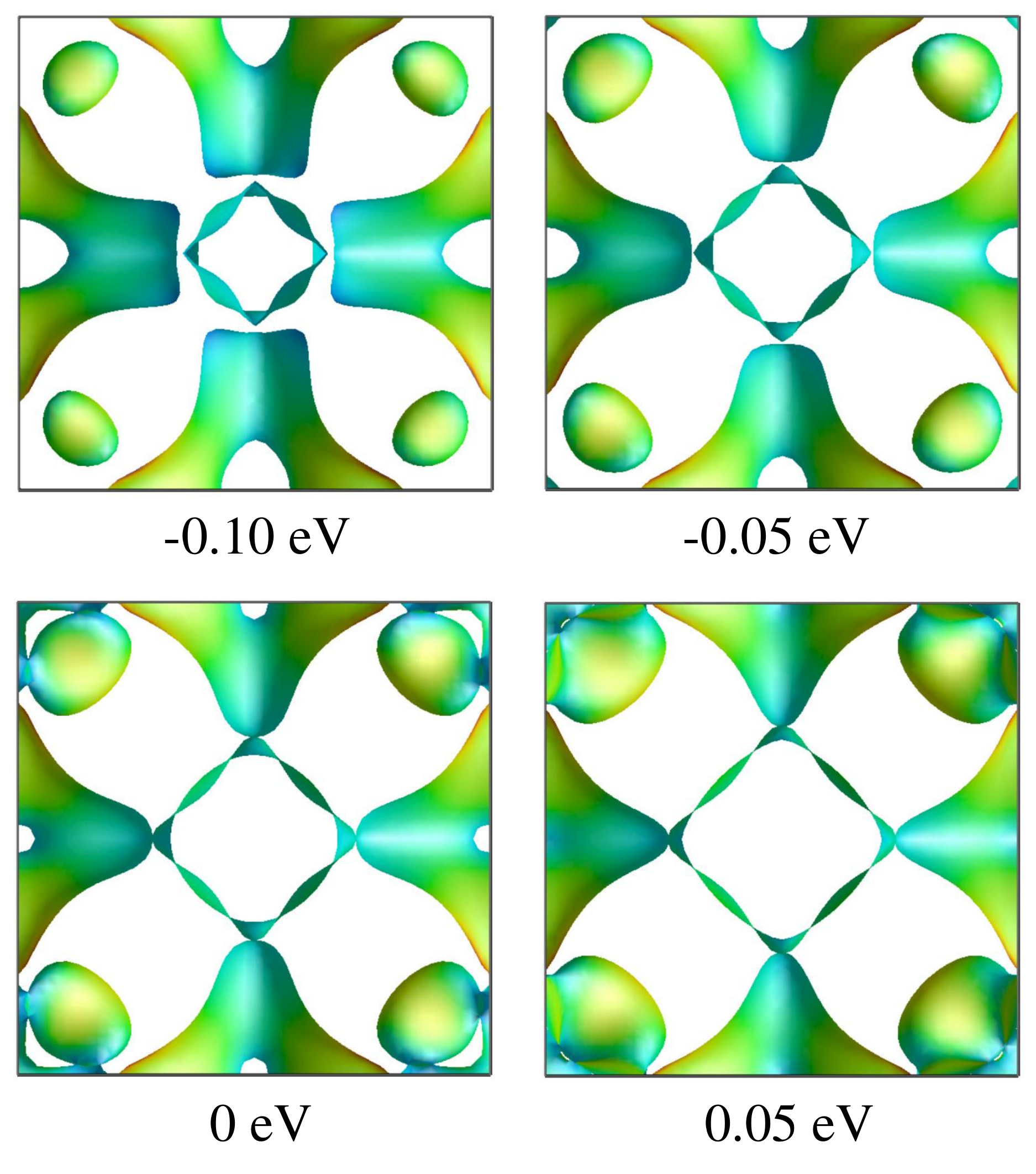}
\caption{$c$-axis view of the Fermi surface as in
Fig. \ref{fermi}, but with shifted Fermi energies. The
corresponding band fillings are 
-0.38 e (-0.10 eV), -0.23 e (-0.05 eV), 0 e (0 eV), and 0.20 e (0.05 eV). }
\label{fermi-e}
\end{figure}

The value of the DOS at the Fermi energy
(with spin-orbit) is $N(E_F)$=4.26 eV$^{-1}$
on a per formula unit basis.
This corresponds to a bare specific
heat $\gamma$=10.0 mJ mol$^{-1}$K$^{-2}$ and bare Pauli susceptibility
$\chi_0({\bf q}=0)$=1.37x10$^{-4}$ emu/mol.
The calculated Drude plasma energies are
$\hbar\Omega_{p,xx}$=2.71 eV and
$\hbar\Omega_{p,zz}$=2.14 eV for the in-plane and $c$-axis directions,
respectively. For isotropic scattering this implies a resistivity
anisotropy of $\sim$1.6, meaning that from a transport point
of view this is a very 3D material.
This is reminiscent of what was found previously by Pickett
for the related compound Na$_2$Sb$_2$Ti$_2$O.
\cite{pickett}

The density of states shows that the O $2p$ derived bands lie between
$\sim$-7 eV and -5 eV, while the Sb $5p$ bands are in the energy
range -4 eV -- -1 eV, all with respect to $E_F$.
Therefore, these shells
are nominally full. Also there are no Ba derived occupied valence
bands. Thus Sb is a trivalent anion in this compound and
Ti is in its trivalent $d^1$ state.
The density of states around $E_F$ is derived primarily
from Ti $d$ states, hybridized with Sb $p$.

The orbital character is illustrated in Fig. \ref{fatbands},
which shows the band structure in a 1 eV range around $E_F$ emphasizing
the character of different $d$-orbitals in a so-called ``fat bands" scheme,
where the bands are shown with symbols
having a size that is a small value to make all bands
visible plus an enhancement of the size proportional to the
projection of given orbital character onto the LAPW sphere.
As may be seen, there are three $d$ orbitals -- $d_{z^2}$, $d_{x^2-y^2}$
and $d_{xy}$ -- that contribute significantly at $E_F$. There is strong
$d_{xy}$ character in this mixture.

The Fermi surface is shown in Fig. \ref{fermi} and the dependence
on band filling is given in Fig. \ref{fermi-e}. There are three
sheets of Fermi surface. The first is a very two dimensional, square electron
cylinder around the zone center. As seen in the fat bands plot, the band
making up this cylinder contains a mixture of the three $d$ orbitals
that participate at $E_F$.
There is a three dimensional
complex shaped electron section around the $\Gamma-Z$ line. This
section has $d_{z^2}$ character.
Finally, there is a large three dimensional hole section around
$X$. Like the 
square cylinder electron section, this part of the
Fermi surface has mixed character derived from the three
active $d$-orbitals, though the dominant character is $d_{z^2}$.
Since the electron count is even, this section compensates
the two electron sections.
This basic structure is preserved as the Fermi level is lowered
within the range corresponding to superconductivity
in Na$_x$Ba$_{1-x}$Ti$_2$Sb$_2$O.
The hole section grows, while the two electron sections shrink.
Importantly, the square cylindrical shape of the 2D section is preserved.

\begin{figure}
\includegraphics[width=0.9\columnwidth]{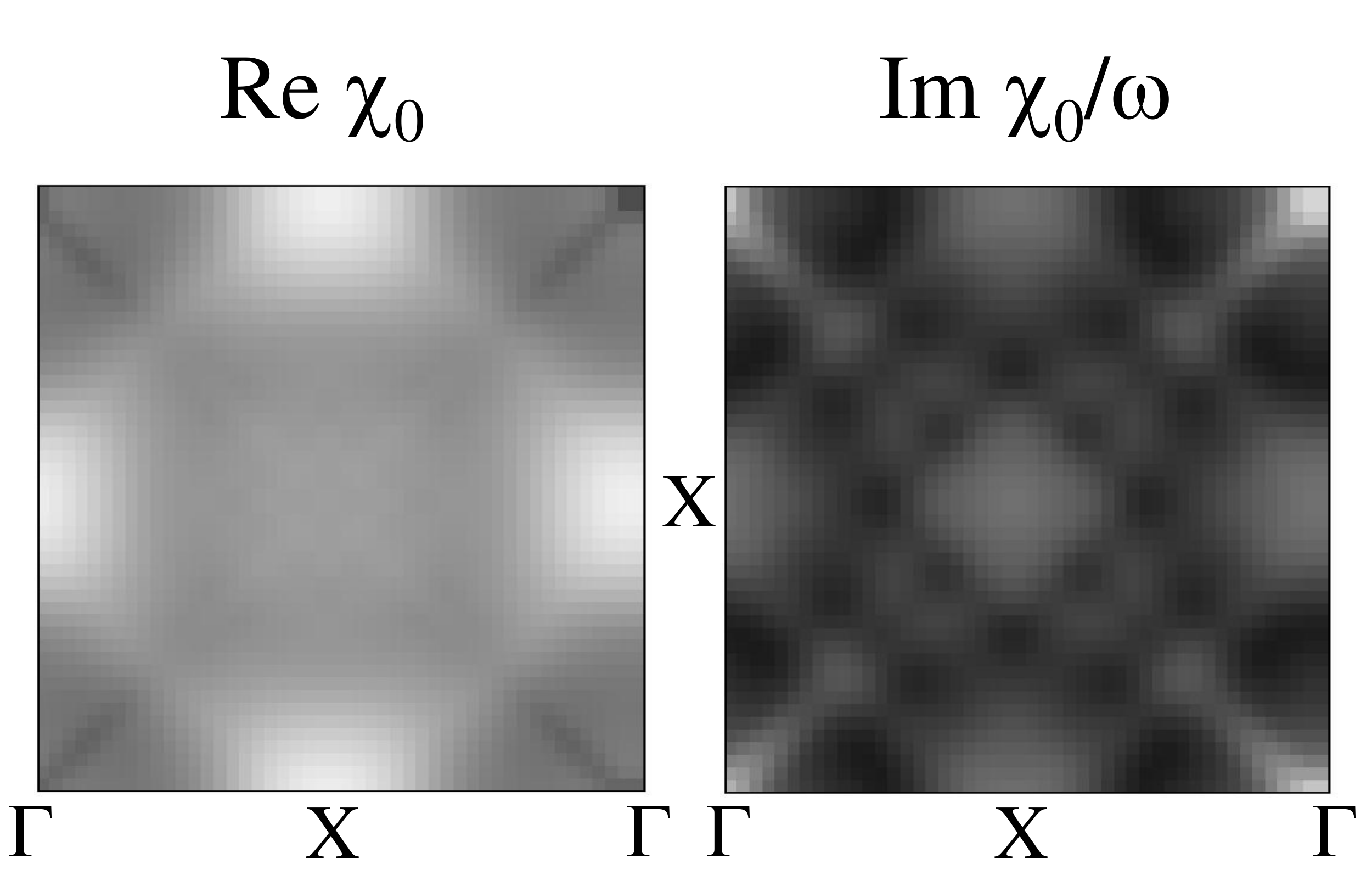}
\caption{Calculated $c$-axis projections of the real and
imaginary parts of $\chi_0$ (see text) with arbitrary units.
Light shading denotes higher values.
}
\label{chi}
\end{figure}

An important aspect of the Fermi surface is that the square
section, as well as the edges of the $X$ centered electron
section are nested along the [1,1] direction. The 2D nesting
vector is approximately (0.24,0.24)(2$\pi$/$a$).
[1,1] is the orientation of the Ti square lattice. In this lattice,
the $d_{xy}$ orbitals are along the Ti-Ti bond directions, and may be
expected to give nested bands. Actually, the nested electron section has
strong $d_{xy}$, but also as mentioned
involves the other two active $d$ orbitals.
In any case,
qualitatively the nesting found should lead to two ridges
in the bare susceptibility $\chi_0$ running along the diagonals in
the zone.
The intersection near the $X$ points would give a
prominent peaks in $\chi_0({\bf q})$ near the $X$ points.
This in fact is the case. Fig. \ref{chi} shows the bare
Lindhard $\chi_0$ calculated with the neglect of the matrix
element approximation, similar to calculations presented
for the Fe-based superconductors. \cite{mazin-spm}
As seen there are diagonal ridges in the real part of $\chi_0$,
with a prominent peak at $X$. This peak will be further increased
by the RPA enhancement,
$\chi({\bf q})$=$\chi_0({\bf q})$/(1-$I({\bf q})\chi_0({\bf q})$)
(here $I({\bf q})$ is written as a Stoner term, but it in correlated
materials it takes the form of the Hubbard parameter $U$).

The real part of $\chi$ governs magnetic ordering as well
as providing the pairing interaction in spin-fluctuation induced
superconductivity.
\cite{berk,moriya}
We discuss it further below.
The imaginary part of $\chi_0$, like the
real part, shows ridges with intersections
near $X$, but the structure is more complex.
This structure in both the
real and imaginary parts of $\chi$,
$i.e.$ intersecting ridges with
peaks at the intersections, is a consequence of the
Fermi surface structure, particularly the square cylinder shape.
This Fermi surface
amounts to the intersection of two 1D sections. The result is similar
to the antiferromagnetic peak in the triplet superconductor Sr$_2$RuO$_4$,
\cite{mazin-214,braden-214}
but different from the Fe-based superconductors where the nesting
arises from matching of electron and hole Fermi surface sections.

We did direct self-consistent
calculations to look for a magnetic ground state. We
find a magnetic instability at the $X$ point.
These calculations were done including spin-orbit
with the moment directed along the $c$-axis.
We find an instability, with small Ti moments inside the LAPW spheres
of 0.2 $\mu_B$. This is much smaller than the $\sim$ 2 $\mu_B$
characteristic of the Fe-based superconductors in similar calculations using
the experimental crystal structure. \cite{mazin-mag}
We also checked for a nearest neighbor antiferromagnetic state, which would
correspond to a zone center instability.
However, we did not find such a state.

\begin{figure}
\includegraphics[width=0.4\columnwidth]{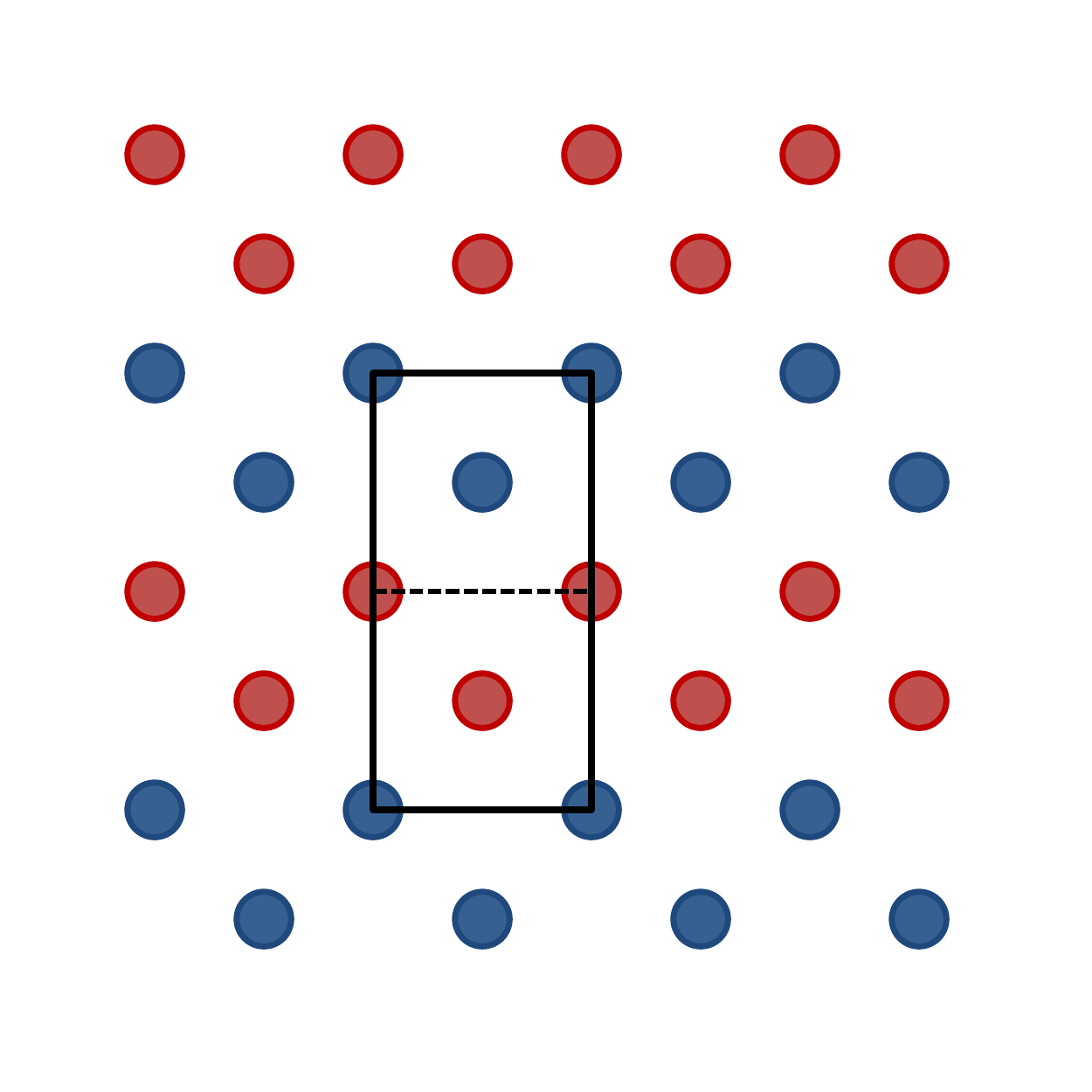}
\includegraphics[width=0.4\columnwidth,angle=90]{ds-mag}
\caption{Magnetic structure of the $X$-point ordering
showing the two degenerate cases in the left and right 
panels. The doubling of the unit cell is shown.
Red and blue denote
up and down spins, respectively. Note the alternation of ferromagnetic and
antiferromagnetic bonds in this double stripe structure.
}
\label{ds}
\end{figure}

The magnetic structure
is shown in Fig. \ref{ds}. This a so-called double stripe structure.
It lowers the symmetry from tetragonal to orthorhombic and
contains both ferromagnetic and antiferromagnetic Ti-Ti bonds,
as shown. This may be expected to result in a spin-dimerization,
with coupling to the lattice, perhaps consistent with neutron
scattering results showing lattice anomalies at $T_s$ in Na$_2$Ti$_2$Sb$_2$O.
\cite{ozawa2}
The actual magnetic structure may be incommensurate, as the moments
are small and as mentioned,
the intersection of the ridges in $\chi({\bf q})$ is off the $X$ point.
Another indication that the actual ground state may be a more complex
incommensurate magnetic
structure  is that in our calculations without imposed
symmetry for the $X$ point instability we find slightly different
moments on the two Ti sites of each spin (the magnetic unit cell contains
four Ti atoms, two spin-up and two spin-down).

In any case, we find a significant coupling of the magnetic order
to the electrons at the Fermi energy. In particular, even though the
moments are small, the DOS near $E_F$ is substantially reduced in the
ordered state. We find $N(E_F)$=3.42 eV$^{-1}$ per formula unit
for the $X$ point ordered state as compared to $N(E_F)$=4.26 eV$^{-1}$
without magnetic order. As in the Fe-based superconductors,
\cite{sebastian}
and in contrast
to cuprates, the antiferromagnetic state near superconductivity is metallic,
both in experiment \cite{doan}
and in the present density functional calculations.

One of the reasons why unconventional superconductivity attracts
so much attention is that it is a rare phenomenon. In particular,
the vast majority of known superconductors are conventional
electron-phonon superconductors, and in particular conventional
electron-phonon superconductivity is known in titanates.
Unlike the Fe-based superconductors, the maximum $T_c$=5.5 K observed in
BaTi$_2$Sb$_2$O is not so high as to preclude conventional superconductivity.
However, the close proximity to an antiferromagnetic state provides
an indication that this may not be the case. In particular, spin-fluctuations
provide a repulsive interaction in a singlet channel, opposite to the
attractive electron-phonon interaction. Therefore the net superconducting
interaction will be $\lambda=\lambda_{ep}-\lambda_{sf}$, where $ep$ and
$sf$ denote electron-phonon and spin-fluctuation contributions, respectively.
As such the wide
coexistence, of magnetic and superconducting order in the composition
dependent phase diagram argues against (but does not fully exclude)
conventional superconductivity. An argument against an unconventional
state would be that scattering due to disorder in the alloy should
suppress an unconventional state. However, the disorder in the present
case is in the Ba layer, and as mentioned Ba does not contribute significantly
to the electronic structure near $E_F$, and furthermore, as mentioned
the bands are rather flat, which means that the
superconducting coherence length will
be relatively short, again reducing the effect of disorder.

\begin{figure}
\includegraphics[width=0.6\columnwidth]{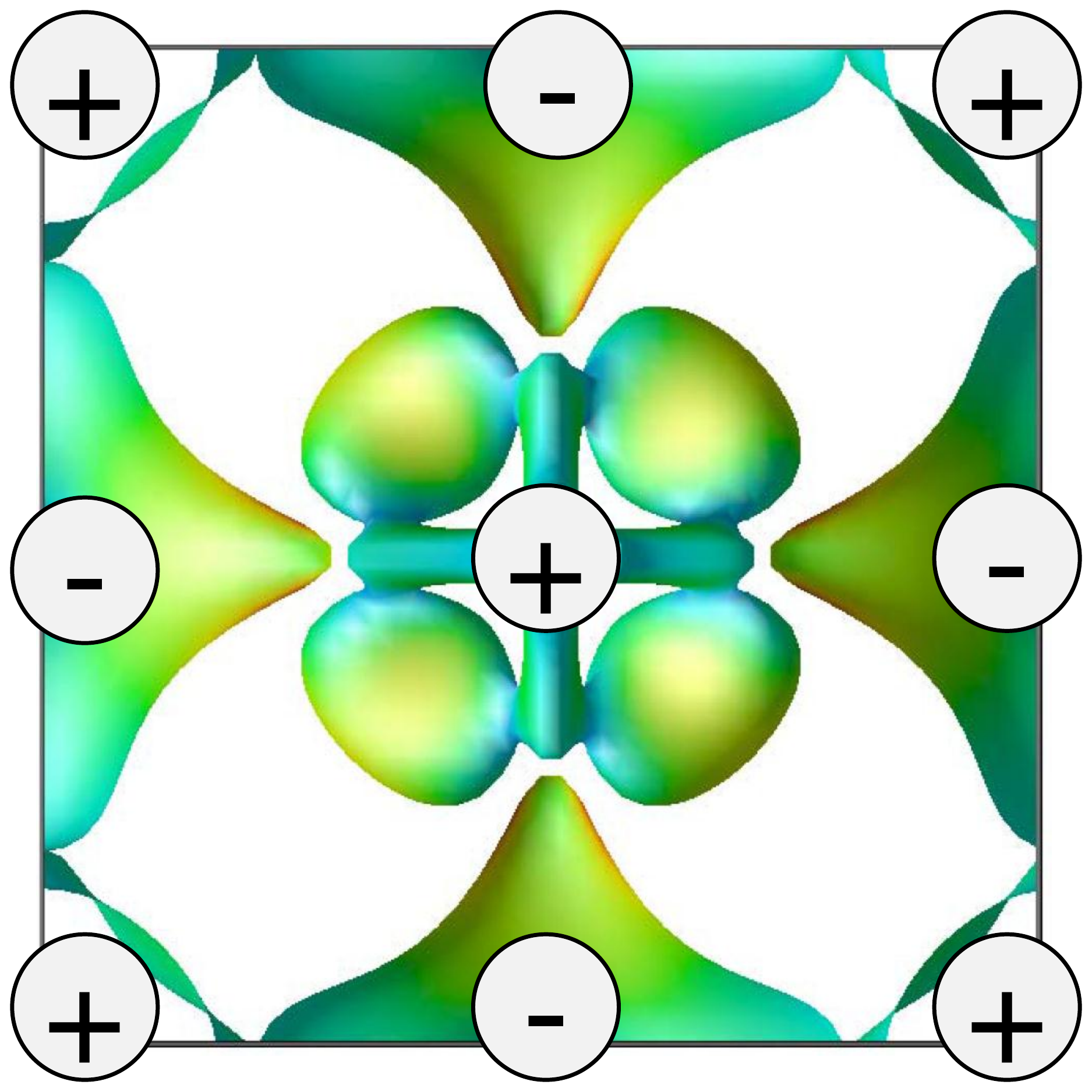}
\caption{Potential sign changing $s$-wave superconducting
state, shown on top of the Fermi surface. Note that
in this figure, $\Gamma$ is at the center, as opposed to
the corners of the plot.
}
\label{fscenter}
\end{figure}

As such, we discuss the possibility of a spin-fluctuation mediated
unconventional superconducting state.
In such a scenario, the pairing interaction is related to the real
part of the actual RPA
enhanced $\chi({\bf q})$ and is repulsive for a singlet state and attractive
for a triplet state. \cite{berk,moriya,scalapino}
As mentioned, in BaTi$_2$Sb$_2$O the susceptibility
is peaked near $X$.
Identification of the superconducting state amounts
to matching the pairing interaction with the Fermi surface
structure, so that opposite sign order parameters occur
on parts connected by peaks in the repulsive (singlet case) interaction
or parts connected by peaks in the attractive (triplet case)
interaction have the same sign order parameter, while maintaining
consistency with the overall triplet or singlet parity.
We find that the best match is obtained for a singlet state
as depicted in Fig. \ref{fscenter}.
In this state the two sections of Fermi surface with similar
orbital character, namely the sqaure section and the large section around
$X$, have opposite sign as expected in such a scenario.

As seen, this is a sign-changing
$s$-wave state, although not the same one as in the Fe-based
superconductors. It consists of a state where the two
electron sections of the Fermi surface have opposite sign order
parameter to the large hole section. The motivation for
this state is that it has sign changes between all pairs of
Fermi surface sections that are separated by (0,$\pi/a$) and
($\pi/a$,0), where the susceptibility is peaked.
Similar to the Fe-based superconductors, the SDW and this superconducting
state are competing instabilities of the Fermi surface.
However, the SDW antiferromagnetism found for BaTi$_2$Sb$_2$O
is much weaker than that in the Fe-based superconductors.
It is unclear what role correlation effects beyond standard
density functional calculations play in these titanates.
The fact that the transition at $T_s$ is a metal-metal
transition in spite of the integral band filling suggests
that they are not so strong as in $e.g.$ the cuprates.
This of course is an expected result considering that the $d$
orbitals in Ti$^{3+}$ ions are more extended than those in 
Cu$^{2+}$. One possible effect of Coulomb correlations will be
to enhance the small moments found here. It will be of interest to
perform neutron diffraction experiments both to confirm whether
the transition is in fact an
SDW and to quantify the magnitude of the moments.
In this regard, Subedi has recently presented electron-phonon
calculations \cite{subedi}
that are consistent with electron-phonon superconductivity
and so either scenario is possible here.

Returning to Na$_2$Ti$_2$Sb$_2$O, that material, \cite{adam}
which has been studied more extensively than BaTi$_2$Sb$_2$O,
has not been reported to show superconductivity, but does have a 
transition, likely of spin density wave or charge density wave character
and a similar phenomenology to BaTi$_2$Sb$_2$O. The transition
in Na$_2$Ti$_2$Sb$_2$O
is at higher temperature, $T_s$$\sim$114 K -- 120 K.
\cite{axtell,liu2,ozawa,ozawa2}
From a structural point of view, Na$_2$Ti$_2$Sb$_2$O
has similar Ti$_2$Sb$_2$O layers to BaTi$_2$Sb$_2$O,
but these are stacked differently along the $c$-axis so that the Sb-Sb
linear chains in the Ba compound are broken up yielding a body
centered tetragonal structure.
The band structure, which as mentioned has similarities to the present
compound, was investigated by Pickett using density functional theory,
and in a tight binding framework by de Biani and co-workers. \cite{biani}
The arsenide, Na$_2$Ti$_2$As$_2$O has a transition at $T_s$$\sim$320 K.
\cite{liu2}
Other related compounds may be (SrF)$_2$Ti$_2$Sb$_2$O, which has
an apparently similar phase transition at $T_s$$\sim$198 K and
(SmO)$_2$Ti$_2$Sb$_2$O with $T_s$$\sim$230 K.
\cite{liu}
BaTi$_2$As$_2$O also has been reported and has similar transition
at $T_s$$\sim$200 K.
Considering the superconductivity of BaTi$_2$Sb$_2$O, it will be of
great interest to study these compounds in more detail and
especially their properties if the SDW can be suppressed, e.g.
by pressure, doping or chemical substitutions.
It will be particularly interesting to study the pressure dependence
in BaTi$_2$Sb$_2$O and related materials.

In conclusion, we find that BaTi$_2$Sb$_2$O has an SDW
instability associated with Fermi surface nesting. The
close proximity of the superconducting phase that emerges
with Na doping to this antiferromagnetic phase suggests
the possibility of unconventional spin-fluctuation mediated
superconductivity. We find that the susceptibility is peaked
near the $X$ point. Matching this structure of the susceptibility
with the Fermi surface suggests
the possibility of a sign-changing $s$-wave
superconducting state of a different nature than the Fe-based
superconductors.

\acknowledgments

This work was supported by the U.S. Department of Energy,
Basic Energy Sciences, Materials Sciences and Engineering Division.
I am grateful to Bernd Lorenz for
helpful discussion and communication of results
and to Igor Mazin and
Douglas Scalapino for useful discussions on spin fluctuations
in relation to superconductivity.

\bibliography{BaTi2Sb2O}

\end{document}